\documentclass{elsart}
\usepackage[dvips]{graphicx}
\usepackage{amssymb}
\usepackage{axodraw}

\begin{document}
\baselineskip 0.6cm
\newcommand{\be}{\begin{equation}}
\newcommand{\ee}{\end{equation}}
\newcommand{\bea}{\begin{eqnarray}}
\newcommand{\eea}{\end{eqnarray}}

\begin{frontmatter}

\hfill HIP-2002-02TH

\title{\bf Single production of 
doubly charged Higgs bosons at hadron colliders}

\author{
J.Maalampi$^{a}$ and
N. Romanenko$^{b}$}

\address{$^a$ Department of Physics, University of Jyv$\ddot{a}$skyl$\ddot{a}$,
 Finland, {\rm and}}

\address{Helsinki Institute of Physics, Helsinki, Finland}
\address{$^b$ Department of Physics at Carleton University,
 Ottawa, Ontario, K1S5B6, Canada}

\thanks[jukka]{jukka.maalampi@phys.jyu.fi}
\thanks[kolya]{nromanen@physics.carleton.ca}

\begin{abstract}

 We reconsider the single production of
 the doubly charged Higgs bosons $\Delta_R^{++}$ and 
$\Delta_L^{++}$  at 
the LHC and TEVATRON in the framework of the left-right
symmetric model and the Higgs triplet model.
 We show that  in the left-right
symmetric model the production of 
$\Delta_R^{++}$
  by Drell-Yan process via $W_R$
exchange may give the dominant contribution. The same channel for
 the production of $\Delta_L^{++}$ in the 
 Higgs triplet model is insignificant.

\bigskip
\end{abstract}

\end{frontmatter}

 Due to the recent convincing
experimental evidences, in particular those on atmospheric and solar
neutrinos \cite{superK,SNO}, the existence of neutrino masses
below the scale 
 $O(1 \, {\rm eV})$  has now become more and more
established. Many theoretical explanations for the lightness of
neutrinos has been suggested and discussed in the literature
\cite{Gun}. Perhaps the most attractive and natural one among
them is the  see-saw mechanism \cite{See-Saw}. In this
mechanism one introduces, in addition to the ordinary Dirac mass
term, Majorana mass terms for the left-handed and 
right-handed neutrinos.
 Neutrinos form a special category among basic
fermions in that they can have, due to their electric
and color neutrality,
such lepton number violating mass terms.

The Majorana mass terms result from the Yukawa couplings of lepton
doublets with  triplet Higgs fields.  The left-handed Higgs
triplet, $\Delta_L = \left( \Delta_L^{0}, \, \Delta_L^{+} , \,
\Delta_L^{++} \right)$,
 that gives rise to the Majorana mass of the left-handed
neutrinos, was first introduced by Gelmini and Roncadelli
\cite{majoron}, while the right-handed Higgs triplet
$\Delta_R = \left( \Delta_R^{0}, \, \Delta_R^{+} , \,
\Delta_R^{++} \right)$
 naturally
appears in the so-called left-right symmetric model \cite{lrm},
where it also takes care of the breaking of the left-right symmetry.

The origin of neutrino masses is quite difficult to figure out
just by looking at the low-energy phenomenology of neutrinos. The
phenomena like neutrino oscillations, leptonic decays of particles
and the neutrinoless double beta decay are quite independent of
the mechanism by which neutrino masses are created. Indirect
information  could, however, be
obtained by studying the physics related to the mass generation
mechanism in high-energy collision process.
Particularly, indirect information on the  see-saw
mechanism can be obtained through the phenomenology
of the triplet Higgs fields.

The phenomenology of Higgs triplets differs quite much from that
of the  \mbox{Standard} Model (SM) Higgs doublets. The most dramatic
difference certainly is the existence of a doubly charged scalar
field ($\Delta^{++}$) within the triplets. In addition,
low-background signals are easy to find due to fact that the
Yukawa couplings of the triplet Higgs bosons violate lepton number
conservation, a symmetry strictly obeyed by the Standard Model
interactions.

By studying the high-energy phenomenology of the triplet Higgs
fields one could get information on the properties of these fields
and on the strengths of their Yukawa couplings \cite{phen},
\cite{lcoll}, \cite{HERA}.
 Several constraints on Yukawa couplings
can be obtained from experimental data \cite{Yuklimits}, the most
stringent limit \cite{GKR} coming from the recent results on
muonium-antimuonium conversion \cite{Willmann}:
\begin{equation}
\frac{\sqrt{h_{ee}\cdot h_{\mu \mu}}}{M_{\Delta^{++}}}
< 0.44 \; {\rm TeV}^{-1} \;  .
\label{yukawalimits}
\end{equation}

The tightening of these limits makes the study of the high-energy
phenomenology of the triplet Higgs bosons at future hadronic
colliders interesting and well motivated.

Most of the previous studies of the production of the doubly
charged Higgs bosons at hadronic colliders have concentrated to
two production mechanisms: single production via $WW$-fusion and
the $\Delta^{++} \Delta^{--}$ pair production  via s-channel photon
and Z-boson exchange. A comparison of these two production
mechanisms at the LHC
environment  was presented  in \cite{HuituMaalampi}. A
more general analysis of various production mechanisms 
of the doubly charged
Higgs bosons 
at future  colliders was done in \cite{GunionLoomisPits}.
This analysis was, however, restricted to the case of the
left-handed Higgs triplet $\Delta_L^{++}$,
 where one can safely neglect the
$WW\Delta$ vertex. This vertex is proportional to the VEV of the
neutral member of the Higgs triplet, which in the case of the
left-handed triplet is very small due to the constraint set by the
 $\rho$ parameter. It was shown that the LHC could probe
the left-handed doubly charged Higgs boson up to a limit
of 950 GeV for
$M_{\Delta^{++}}$ .
 The LHC discovery potential in the case of
the process $p\,p \rightarrow e^-\, e^-\,
\mu^-\, \mu^-$ was estimated in \cite{ADatta} to be
$M_{\Delta_L^{++}} < 850 \,$ GeV. It should be noted that both of
these numerical results were obtained with the assumption that
$\Delta_L^{++}$ decays to leptons only.

In this paper we shall consider some new aspects, not addressed 
in the previous studies,
of the production
of the doubly charged Higgs bosons at hadron colliders LHC and
TEVATRON. In
particular, we will investigate the single $\Delta^{++}$
production via the reaction $p\,p \rightarrow \Delta^{++} q\,q
$ (see Fig. 1 a and b).
This $WW$ fusion process is experimentally interesting due to its
clear bilepton signature  from the decay  of the doubly
charged Higgs boson. We also study the Drell-Yan-type
 production  of
a single $\Delta^{++}$ via the process  $p\,p \rightarrow \Delta
\, W \rightarrow \Delta^{++} q\,q $ (see Fig. 1 c and d). As we
will see below, this process can be the dominant production
mechanism for a wide range of parameter values in the case of the
 $\Delta_R^{++}$.
 Both of the processes go through the $WW\Delta$ vertex. 
  The production threshold
in the single production processes 
is lower than that in the pair
production, and therefore these processes are particularly
interesting phenomenologically as they probe
 a larger mass range of
$\Delta^{++}$.

\begin{figure}
{
\begin{center}
\unitlength=1.0 pt
\SetScale{1.0}
\SetWidth{0.7}      
\scriptsize    
{} \qquad\allowbreak
\begin{picture}(95,79)(0,0)
\Text(15.0,70.0)[r]{$u$}
\ArrowLine(16.0,70.0)(58.0,70.0)
\Text(80.0,70.0)[l]{$d$}
\ArrowLine(58.0,70.0)(79.0,70.0)
\Text(54.0,60.0)[r]{$W^+$}
\DashArrowLine(58.0,70.0)(58.0,50.0){3.0}
\Text(80.0,50.0)[l]{$\Delta^{++}$}
\DashArrowLine(58.0,50.0)(79.0,50.0){1.0}
\Text(54.0,40.0)[r]{$W^+$}
\DashArrowLine(58.0,30.0)(58.0,50.0){3.0}
\Text(15.0,30.0)[r]{$u$}
\ArrowLine(16.0,30.0)(58.0,30.0)
\Text(80.0,30.0)[l]{$d$}
\ArrowLine(58.0,30.0)(79.0,30.0)
\Text(47,0)[b] {(a)}
\end{picture} \
\unitlength=1.0 pt
\SetScale{1.0}
\SetWidth{0.7}      
\scriptsize    
{} \qquad\allowbreak
\begin{picture}(95,79)(0,0)
\Text(15.0,70.0)[r]{$u$}
\ArrowLine(16.0,70.0)(58.0,70.0)
\Text(80.0,70.0)[l]{$d$}
\ArrowLine(58.0,70.0)(79.0,70.0)
\Text(54.0,60.0)[r]{$W^+$}
\DashArrowLine(58.0,70.0)(58.0,50.0){3.0}
\Text(80.0,50.0)[l]{$\Delta^{++}$}
\DashArrowLine(58.0,50.0)(79.0,50.0){1.0}
\Text(54.0,40.0)[r]{$W^+$}
\DashArrowLine(58.0,30.0)(58.0,50.0){3.0}
\Text(15.0,30.0)[r]{$\bar{d}$}
\ArrowLine(58.0,30.0)(16.0,30.0)
\Text(80.0,30.0)[l]{$\bar{u}$}
\ArrowLine(79.0,30.0)(58.0,30.0)
\Text(47,0)[b] {(b)}
\end{picture} \
\newline
\begin{picture}(95,79)(0,0)
\Text(15.0,70.0)[r]{$u$}
\ArrowLine(16.0,70.0)(37.0,60.0)
\Text(15.0,50.0)[r]{$\bar{d}$}
\ArrowLine(37.0,60.0)(16.0,50.0)
\Text(47.0,64.0)[b]{$W^+$}
\DashArrowLine(37.0,60.0)(58.0,60.0){3.0}
\Text(80.0,70.0)[l]{$\Delta^{++}$}
\DashArrowLine(58.0,60.0)(79.0,70.0){1.0}
\Text(80.0,50.0)[l]{$W^-$}
\DashArrowLine(58.0,60.0)(79.0,50.0){3.0}
\Text(47,0)[b] {(c)}
\end{picture} \
\unitlength=1.0 pt
\SetScale{1.0}
\SetWidth{0.7}      
\scriptsize    
{} \qquad\allowbreak
\begin{picture}(95,79)(0,0)
\Text(15.0,70.0)[r]{$u$}
\ArrowLine(16.0,70.0)(37.0,60.0)
\Text(15.0,50.0)[r]{$\bar{d}$}
\ArrowLine(37.0,60.0)(16.0,50.0)
\Text(47.0,64.0)[b]{$W^+$}
\DashArrowLine(37.0,60.0)(58.0,60.0){3.0}
\Text(80.0,70.0)[l]{$\Delta^{++}$}
\DashArrowLine(58.0,60.0)(79.0,70.0){1.0}
\Text(54.0,50.0)[r]{$W^-$}
\DashArrowLine(58.0,60.0)(58.0,40.0){3.0}
\Text(80.0,50.0)[l]{$d$}
\ArrowLine(58.0,40.0)(79.0,50.0)
\Text(80.0,30.0)[l]{$\bar{u}$}
\ArrowLine(79.0,30.0)(58.0,40.0)
\Text(47,0)[b] {(d)}
\end{picture} \
{} \qquad\allowbreak
\end{center}}
\caption{ The Feynman diagrams for the single $\Delta^{++}$
production via the  $WW$ fusion ((a) and (b)),
and via the Drell-Yan
process  ((c) and (d)).}
\end{figure}
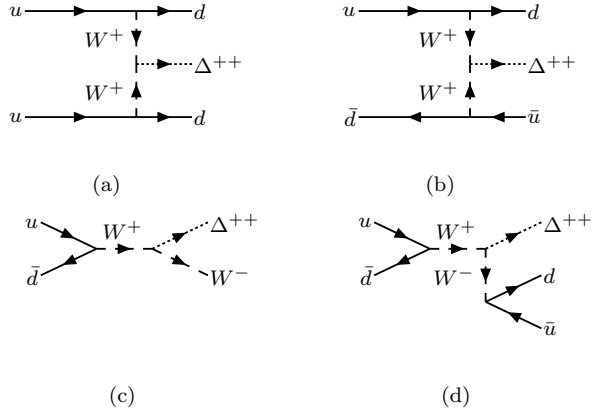 
We have studied the $WW$ fusion and Drell-Yan processes in two
models, in the left-right symmetric model (LRM) and in the
Higgs Triplet model (HTM)
where  an additional Higgs triplet is added to the SM.
 A detailed
description of the LRM  is found in \cite{lrm}, whereas the HTM
 is discussed in \cite{godbole}.

   The have evaluated the production cross sections
by using the CompHEP package \cite{comphep}. The Feynman rules for
the HTM were checked with the help of the LanHEP
package \cite{lanhep}. As usual, we have calculated cross sections
with two initial quarks and appropriately  convoluted them with
the quark structure functions:
\begin{equation}
\sigma_{p,(anti)p \rightarrow final \; state}=
\Sigma_{i,j}
\int dx_1 dx_2 f_i(x_1) \cdot f_j(x_2)
\cdot \hat{\sigma}_{ij}(x_1P_1, x_2P_2)
\end{equation}
where $i,j$ enumerate quarks inside protons (antiprotons),
$x_{1,2}$ denote the fraction of incident momenta $P_{1,2}$
carried by the quarks and $\hat{\sigma}$ denotes the appropriate
quark cross section. For our calculations we have used the MRS
structure functions implemented by the CompHEP. For the final state quark
jets  we have applied  an angular cut of $ |\cos(\theta)|< 0.9$ and
an energy cut of $E> 10$ GeV.

The production cross sections of the right-handed
 and left-handed  Higgs triplet
boson at LHC energy ($\sqrt{s}=14$ TeV) 
are presented in Fig. 2. We show separately
cross sections arising from the $WW$ fusion and from the Drell-Yan
process.
 For the case of the LRM
(see Fig. 2a) one finds that at relatively small values of the mass 
of the doubly charged
Higgs boson the Drell-Yan process via $W_R$ yields the dominant
contribution. The cross section depends on the $W_R$ mass, and the
greater this mass is, the more heavy $\Delta^{++}$ can be probed, 
assuming that the Drell-Yan process is the dominant production mechanism, as it is
for the most probable values of the masses of $W_R$
and $\Delta_R^{++}$. 

\begin{figure}[t]
\setcounter{figure}{1}
\begin{center}
\vspace{1cm}
\includegraphics[width=6.5cm, height=8cm]{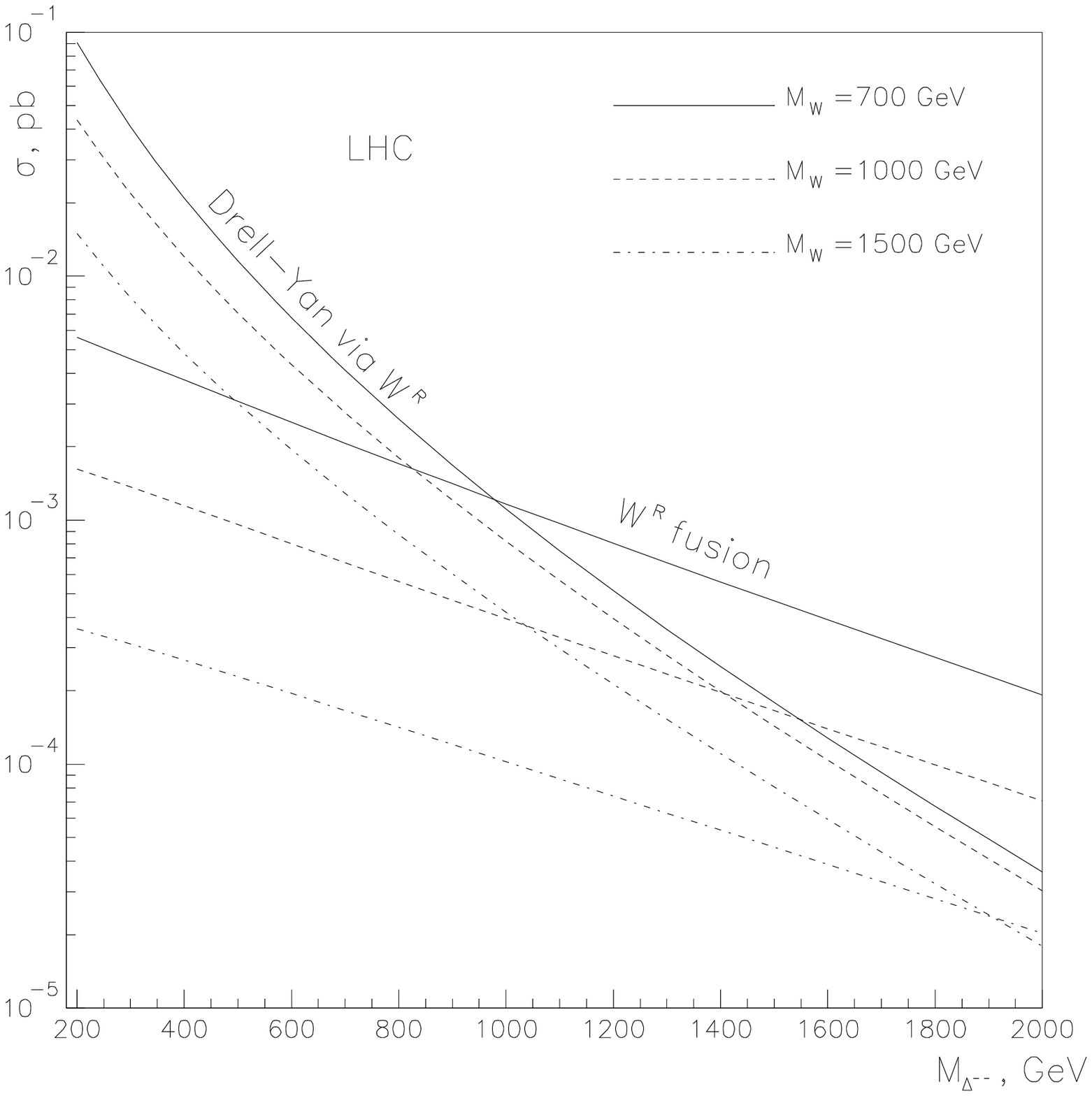}
\includegraphics[width=6.5cm, height=8cm]{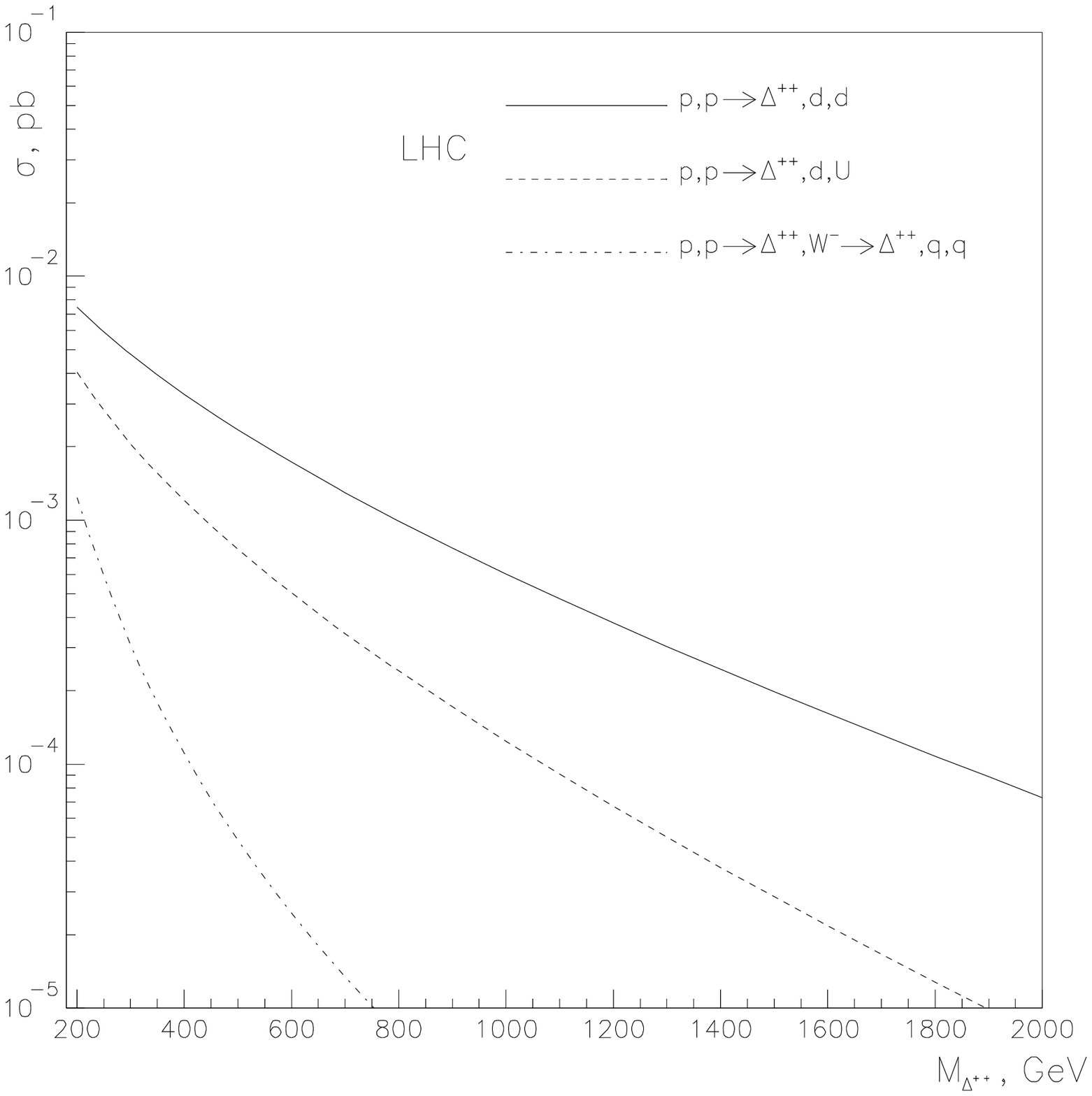}
\centerline{\bf (a) \hspace*{6cm} (b)} 
\caption{The production cross sections of
$\Delta_R^{++}$ (a) and
$\Delta_L^{++}$ (b)
at the LHC
as a function of the mass
$M_{\Delta^{++}}$.
In  (a) the solid line corresponds
to $M_{W_R}=700$ GeV, the dashed line 
to $M_{W_R}=1000$ GeV, the dash-dotted line 
to $M_{W_R}=1500$ GeV. In  (b)
the solid line corresponds
to  $W_L$ fusion with two $d$-jets
in the final state ($p\,p \, \rightarrow \Delta^{++}\,d\,d$),
the dashed line corresponds
 $W_L$ fusion with  $d$- and
$\bar{u}$-jets
in the final state ($p \, p \, \rightarrow \Delta^{++}\,d\,\bar{u}$),
the dash-dotted line corresponds
to the Drell-Yan process with s-channel
 $W_L$ exchange ($p \,p \, \rightarrow \Delta^{++}\,W_L
\, \rightarrow \Delta^{++}\,q\,q $).
}
\label{fig2}

\end{center}
\end{figure}

 However, for the left-handed triplet Higgs boson (see Fig. 2b) situation
is completely different. Here the $W_LW_L$-fusion is the main production channel
for all reasonable values of the doubly charged Higgs boson
mass. The explanation of this fact is the essential difference
between $W_L$ and $W_R$ masses
($M_{W_L}=80$ GeV, $M_{W_R} >700$ GeV \cite{PDG} ).

  In Figure 3 we show the cross sections of the single
$\Delta^{++}$ production at the Tevatron. One can conclude that
for the anticipated  annual luminosity  of the Tevatron,
$\sim 15 \rm{fb}^{-1}$,
 \cite{Sultansoy} the left-handed
doubly charged Higgs bosons are 
unobservable. On the other hand, several the right-handed triplet boson $\Delta_R^{++}$
 events were possible to detect if $M_{W_R}$ 
is near 100 GeV, the
$W_RW_R$--fusion giving the dominant contribution.

\begin{figure}[t]
\setcounter{figure}{2}
\begin{center}
\vspace{1cm}
\includegraphics[width=6.5cm, height=8cm]{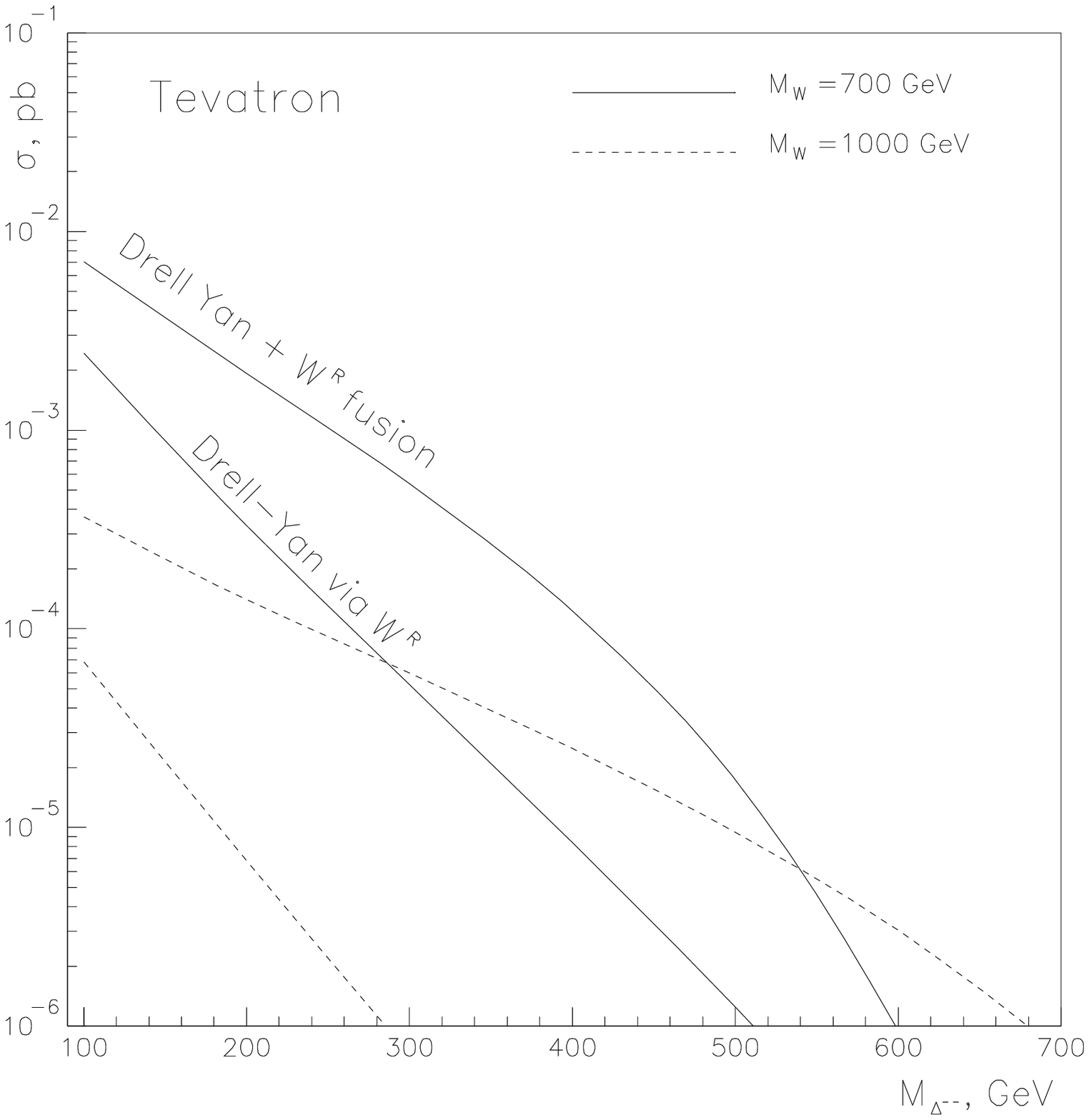}
\includegraphics[width=6.5cm, height=8cm]{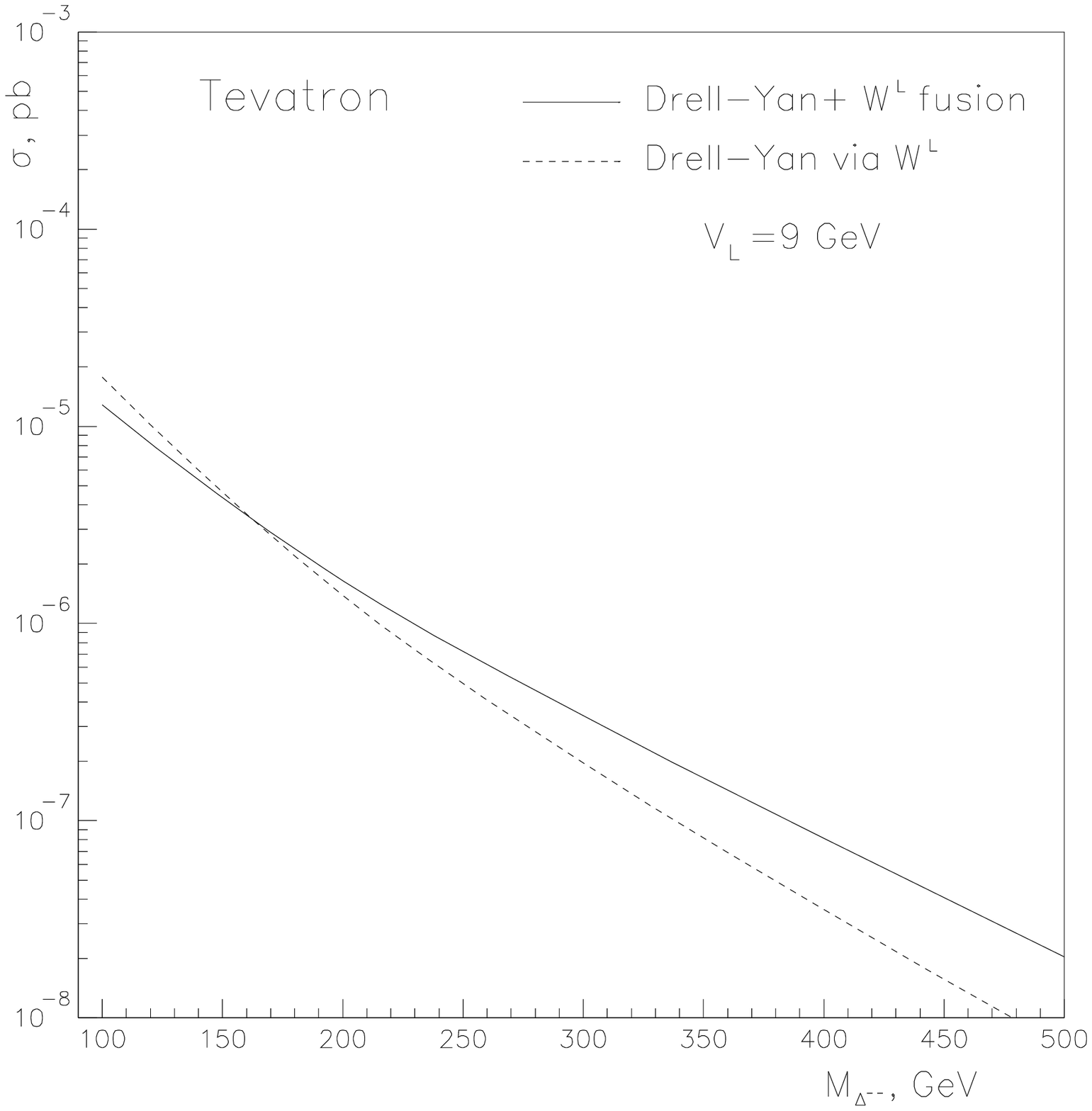}
\centerline{\bf (a) \hspace*{6cm} (b)} \label{fig3}
\caption{ The production cross sections of
$\Delta_R^{++}$ (a)
  and
$\Delta_L^{++}$ (b)
at the TEVATRON as a function of the mass
 $M_{\Delta^{++}}$.
In  (a) the solid line corresponds
to $M_{W_R}=700$ GeV, the dashed line c
to $M_{W_R}=1000$ GeV.
In  (b)
the solid line corresponds
to the process of  $W_L$ fusion with  $d$- and
$\bar{u}$-jets
in the final state ($p\,\bar{p} \, \rightarrow \Delta^{++}\,d\,\bar{u}$),
dashed line corresponds
to the Drell-Yan process with s-channel
 $W_L$ exchange ($p\,\bar{p} \, \rightarrow \Delta^{++}\,W_L
\, \rightarrow \Delta^{++}\,q\,q   $).
}
\end{center}
\end{figure}

  The best signature for the observation of the doubly charged
Higgs bosons comes from the decay mode $\Delta^{++}\rightarrow l^+ l^+$.
This mode may be dominant, depending on the mass splittings between scalars within the triplets and the vacuum expectation value of Higgs triplet
\cite{phen}.
The same-sign two-lepton background from the SM processes is always associated with at least two neutrinos
due to  lepton number conservation,
while the signal process has  very little missing energy.

\begin{figure}[t]
\setcounter{figure}{3}
\begin{center}
\vspace{1cm}
\includegraphics[width=6cm, height=6.5cm,angle=0]{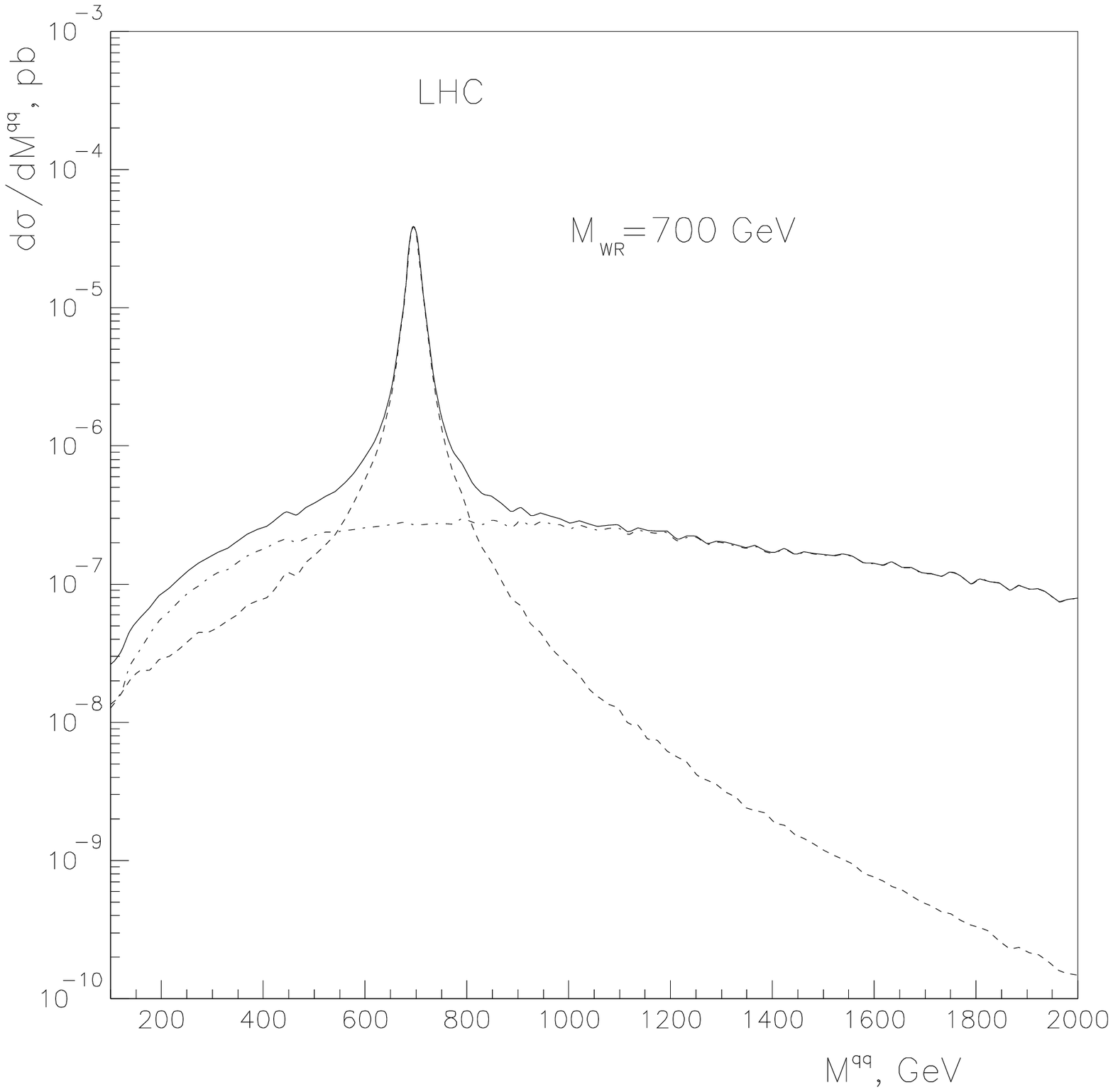}
\includegraphics[width=6cm, height=6.5cm, angle=0]{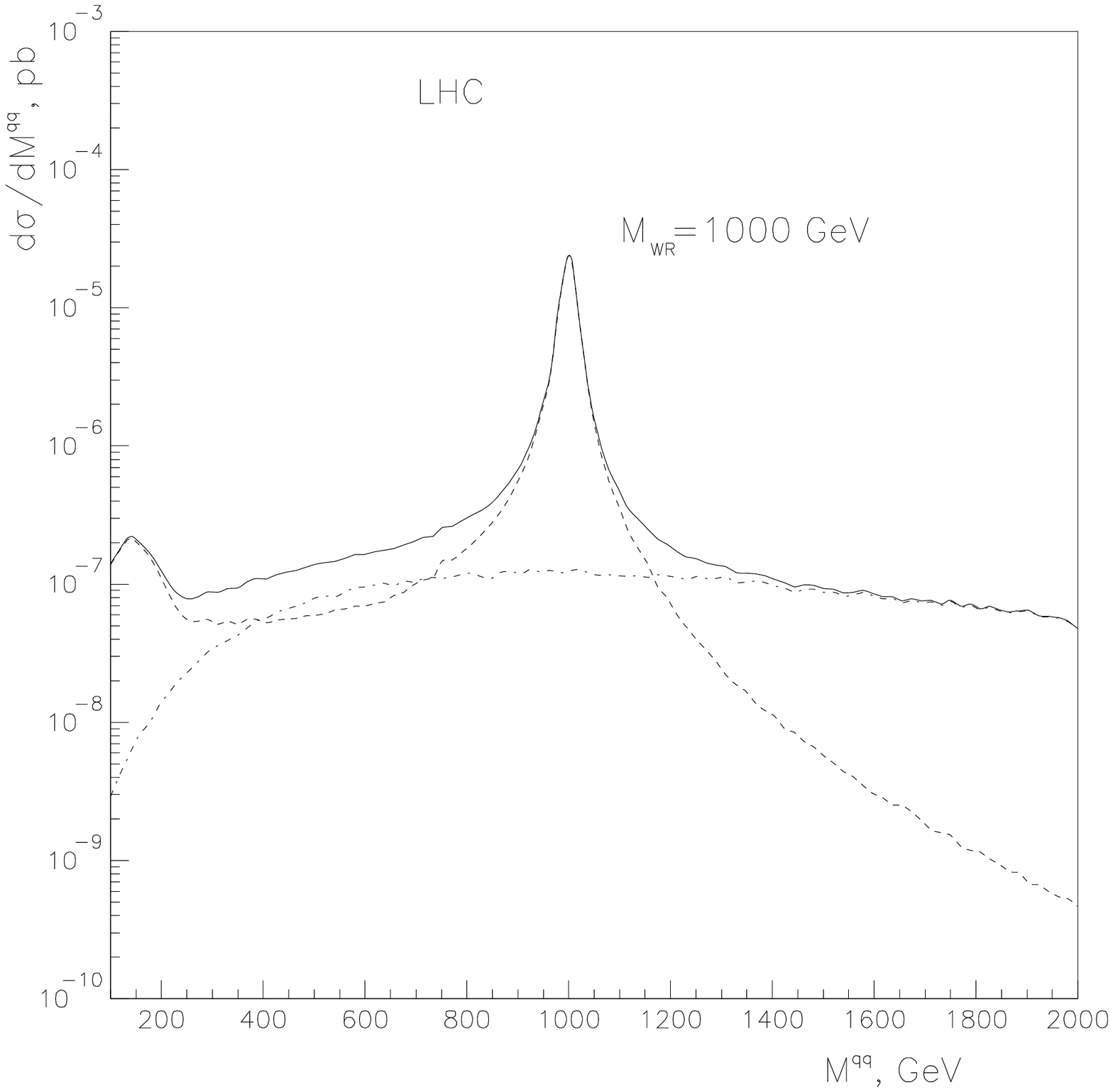}
\centerline{\bf (a) \hspace*{6cm} (b)}
\includegraphics[width=6cm, height=6.5cm, angle=0]{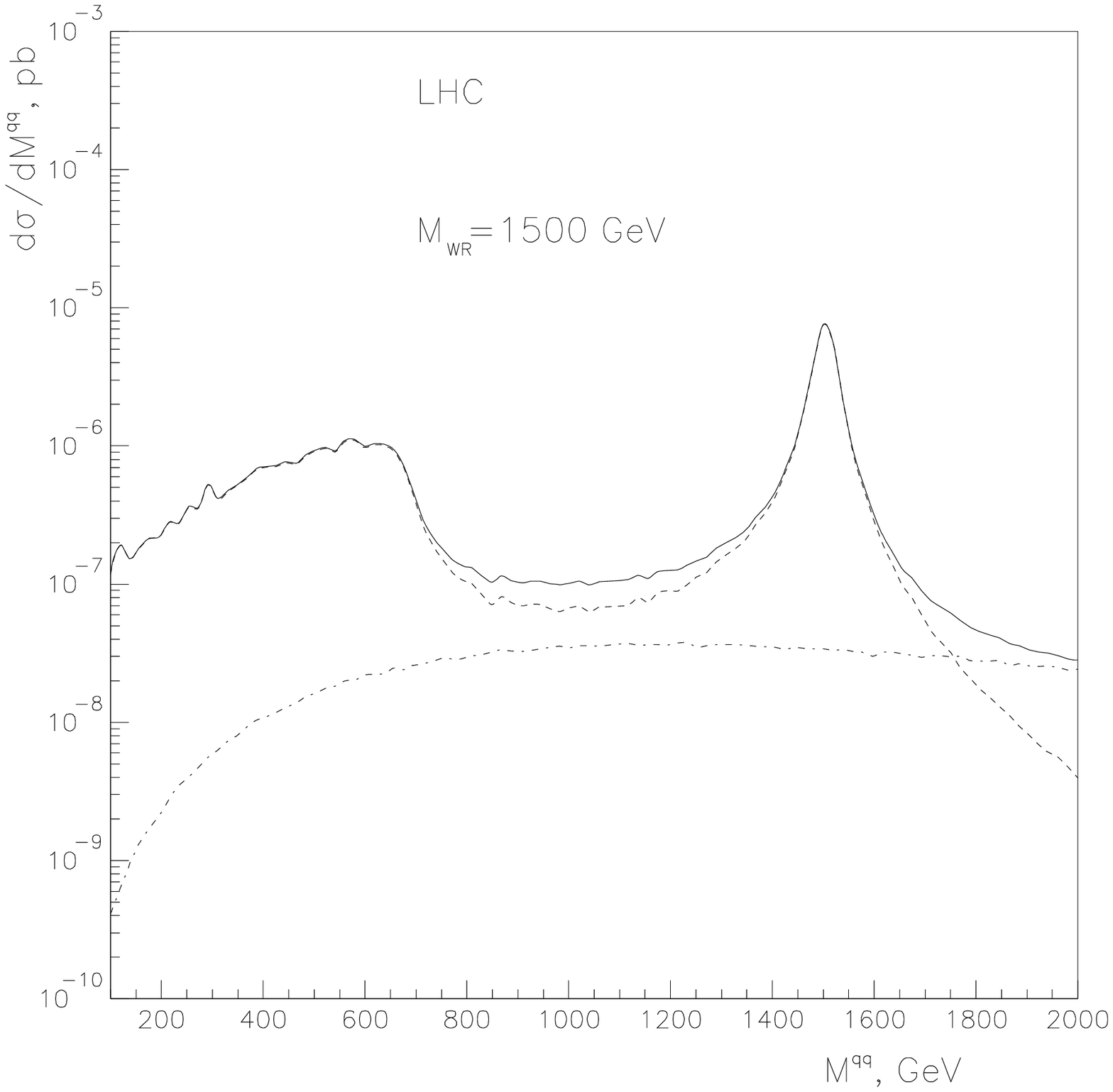}
\centerline{\bf (c)}
\label{fig4}
\caption{ The invariant mass distribution
of the final quarks
 at the LHC  for different values of $M_{W_R}$.
The mass $M_{\Delta^{++}}= 800$ GeV.
 The solid line corresponds to the total cross section,
the dashed line to the Drell-Yan contribution, the dashed-dotted line
 to the contribution of fusion. (The small structures of the curves are due
 to numerics).}
\end{center}
\end{figure}

 In order to distinguish the Drell-Yan and fusion processes
 in the case of the \mbox{LRM,} we have studied
the dependence of the cross section on the opening angle between the
two outgoing like-sign leptons. For the  fusion channel the
cross section occurs to be more forwarded peaked than for the 
Drell-Yan process. However, it turns out that the shape of the distribution
strongly depends on the mass of the doubly charged Higgs boson and makes it
practically impossible to disinguish between the two channels.

Nevertheless, these two channels can be separated due to completely different
invariant mass distributions of the final state quarks. One can see from Fig. 1,
that in the Drell-Yan case the corresponding outgoing quarks
are mostly produced by decays of real $W_R$'s and hence
their  invariant mass distribution 
should have peak at $M^{qq}= M_{W_R}$. It is not true for the fusion channel;
the corresponding distribution should not have this kind of peak.
In Figure 4 we show the $M^{qq}$ distributions for the both channels.
The doubly charged Higgs boson mass is chosen to be 800 GeV,
while
$M_{W_R}$ mass is set to 700 GeV { (a)}, 1000 GeV 
{ (b)} and 1500 GeV  { (c)}.
Of course, the actual data will contain both channels, and we show this
actual   $M^{qq}$ distrubution by solid curve in each of three cases of Figure 4,
the distribution for the Drell-Yan channel is shown by a dashed line,
for the fusion channel  by a dashed-dotted line. 
As a matter of fact, the actual $M^{qq}$ distribution is  to a good extent
just a sum of Drell-Yan and fusion distributions, since for the case of fusion
quark-antiquark final states are suppressed
and may be safely neglected  in comparison with
quark-quark final states, while the latter are totally absent
in the Drell-Yan channel. According to this, if one throws away
the suppressed fusion final states,
 there cannot be any interference between the
two  channels. 

 One can notice an important threshold behaviour in
the $M^{qq}$ distribution of the Drell-Yan
 channel (at $M^{qq}=200$ GeV in Fig 4b and 
at $M^{qq}=700$ GeV in Fig 4c):
 when $M^{qq} + M_{\Delta^{--}} \ge M_{W_R}$,
 the differential cross section turns down.
 This happens due to an s-channel $W_R$ propagator suppression:
 due to structure functions
  the value of $s$ of the incoming quarks is not fixed and it may
  include the $W_R$--propagator  resonance.
 However, as soon as $M^{qq}$ exceeds the abovementioned limit, the
 $W_R$ propagator goes off-shell.
 We used  
 $\Gamma _{W_R}=  \Gamma _{W_L} \cdot M_{W_R}/ M_{W_L}$
 for the $W_R$ width. 
The main contribution to cross sections of Fig. 4
comes from real $W_R$'s of the Drell-Yan channel, with a subsequent decay
of $W_R$ to quark-antiquark pairs.

\begin{figure}[t]
\setcounter{figure}{4}
\begin{center}
\vspace{1cm}
\includegraphics[width=8cm, height=6.5cm,angle=-90]{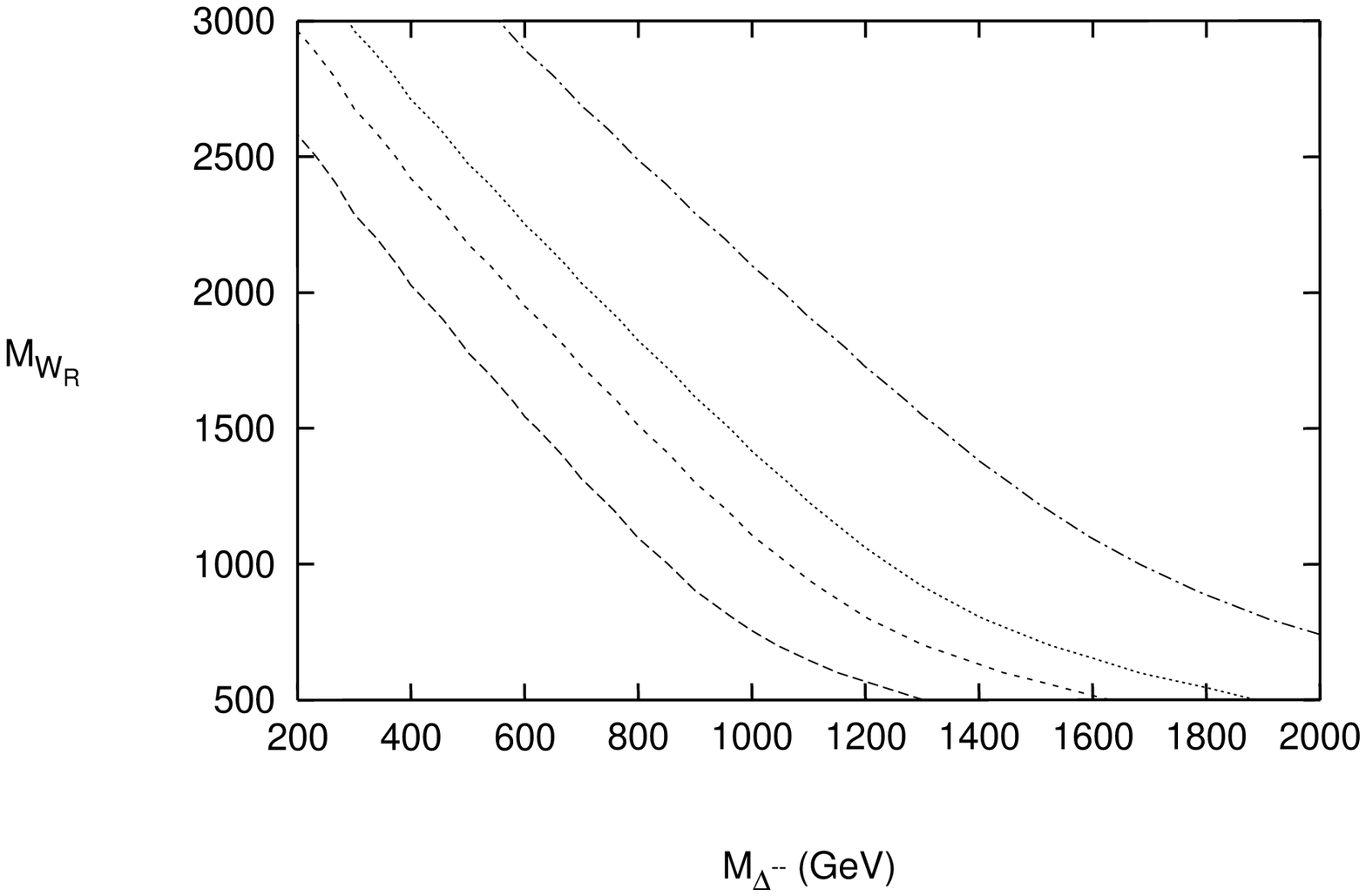}
\includegraphics[width=8cm, height=6.5cm, angle=-90]{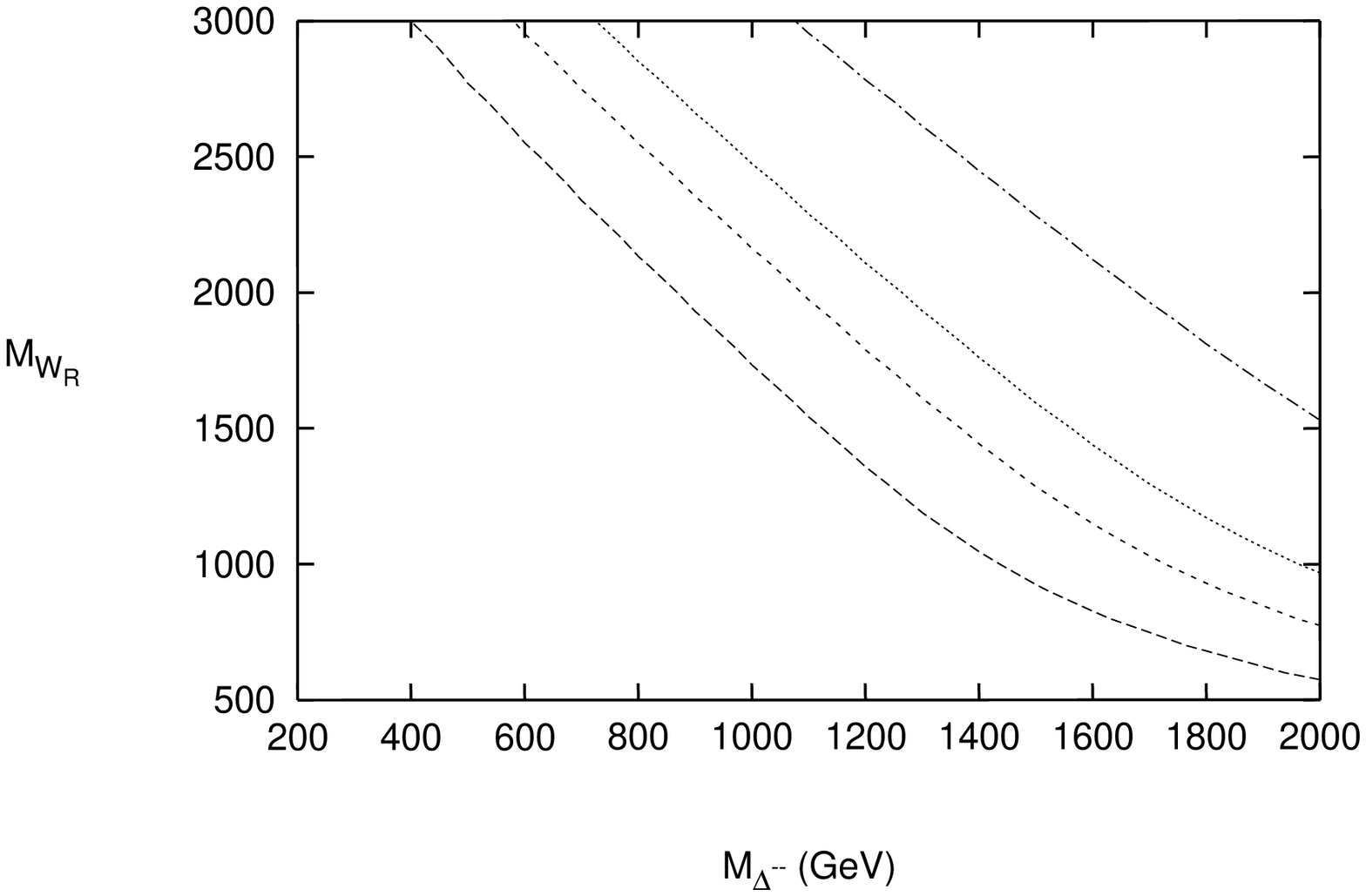}
\centerline{\bf (a) \hspace*{6cm} (b)}
\label{fig5}
\caption{  The LHC discovery limits  of $\Delta_R^{++}$
in the LRM in the $\left( M_{W_R} \; M_{\Delta^{++}} \right)$-
plane at 1, 3, 5 and 10
events levels. The case {\ (a)} corresponds to dominantly bosonic
 decays, case { (b)} corresponds to
leptonic decay channels only.}
\end{center}
\end{figure}

 In Figure 5 we present the discovery limits of
the single production of $\Delta^{++}$
at LHC in the LRM. The contours are drawn in the
$\left( M_{W_R} \, \,M_{\Delta^{++}} \right)$- plane, 
and they correspond to the
1, 3, 5 and 10 events for the  luminosity of  100 $fb^{-1}$, with the
$0.9$ detection efficiency of the each final state muon (lepton).
We have considered two light-quark jets
in the final state
(i.e. $u,d,s,c$ and corresponding antiquarks),
and the branching ratio
of $W_R$ decay into  light quark-antiquark pair
was estimated to be $0.65$, i. e.
 we have neglected the bosonic decay modes for
$W_R$. The
number of  observed events in this case strongly
 \mbox{depends}
 on the branching ratio of $\Delta^{++} \rightarrow \mu^{+}\, \mu^{+}$. 
For $M_{\Delta^{++}} > 2  M_{W_R}$
the decay channel $\Delta^{++}\rightarrow
W_R \, W_R$ is opened. However, even before that other
 bosonic decay channels, e.g.  $\Delta^{++} \rightarrow
\Delta^+ \, W^+_R$ and
 $\Delta^{++} \rightarrow
\Delta^+ \, \Delta^0$ , may become possible. 
Their decay widths strongly depend
 upon the at present unknown mass spectrum of the triplet scalars
and Higgs self-couplings.
In order to take this ignorance
into account 
we take the branching ratio
\begin{equation}
Br_{\mu^+ \mu^+} \equiv
\Gamma(\Delta^{++} \rightarrow \mu^{+}\, \mu^{+})/
\Gamma(\Delta^{++} \rightarrow all)
\end{equation}
 as a phenomenological parameter in our estimates.
We assume the maximal natural value of this
branching ratio to be $1/3$, which corresponds to the case where
the doubly charged
Higgs boson decays always to leptons with the equal flavour
diagonal Yukawa couplings
for all three families.
The plots in Fig. 5 a correspond to the
$Br_{\mu^+ \mu^+}=0.06$,
 and the plots in Fig. 4 b -
as the most  optimistic case -
corresponds to
$Br_{\mu^+ \mu^+}=1/3$.

\begin{table}[t]
\caption{ The 95\% probability mass discovery limits
(given in GeV) of doubly charged
Higgs bosons  in LRM.
Results are shown  for
$Br(\Delta^{++}\rightarrow \mu^+ \, \mu^+)= 0.06$,
0.1, 0.15, 0.2, 0.25, 0.3, 0.33.
Coloumns correspond to the
$M_{W_R}= 1000$, 1500,  2000 GeV.
  In all cases we
assume an integrated luminosity of ${\rm  L}=100$ fb$^{-1}$.
}
\label{table}
\vspace{0.4cm}
\begin{center}
\begin{tabular}{lccc}
\hline
\hline
 $Br.ratio $  &
   {$M_{W_R}=1000$ } &
  {$M_{W_R}$=1500 } &
   {$M_{W_R}$=2000 } \\
  { $ _{
  (\Delta^{++}\rightarrow \mu^+ \, \mu^+)}$}    &
    $M_{\Delta^{++}} $ &
     $M_{\Delta^{++}} $     &  $M_{\Delta^{++}} $    \\
\hline
0.06  & 1250 & 960   & 700   \\
0.1   & 1400 & 1100  & 870   \\
0.15  & 1600 & 1250  & 990   \\
0.2   & 1730 & 1360  & 1080  \\
0.25  & 1820 & 1450  & 1170  \\
0.3   & 1900 & 1520  & 1200  \\
0.33  & 1970 & 1560  & 1250   \\
\hline \hline
\end{tabular}
\end{center}
\end{table}

In Table 1 we present the 95\% C.L. discovery limits
of the $M_{\Delta^{++}}$  for the LHC
in the LRM. We choose values of the
$W_R$ boson mass to be 1000, 1500, or 2000 GeV
and calculate the discovery limits for the different values
of the branching ratio $Br_{\mu^+ \mu^+}$.
As can be seen from the Table, even for a relatively heavy
right-handed W-boson the discovery limit for
$\Delta^{++}$ remains quite high.

In conclusion, the single production of the doubly charged Higgs bosons
via the processes of Fig. 1
offers a way to improve  significantly 
the discovery limits of $\Delta_R^{++}$
for a wide range of $W_R$ masses.
They also probe the $WW\Delta$ vertex which is
inevitable for the triplet Higgs interactions.
However, we showed that
for the TEVATRON energies it is impossible to detect the effects
of theses vertices, while
 at the
 LHC energies
 they really dominate the 
$\Delta_R^{++}$ production.
The analysis of final quarks $M_{qq}$
distribution may reveal the effects
of Drell-Yan channel contribution due to
characteristic peak behaviuor at  
$M^{qq}= M_{W_R}$.
The discovery of this peak would be an indication
of the existence
doubly charged Higgs boson
$\Delta_R^{++}$.

\section*{Acknowledgements}
 Work is supported by the Academy of Finland under the contract  no. 42328
and by RFFI  grant 01-02-17152
(Russian Fund of Fundamental \mbox{Investigations}).

\end{document}